%Paper: hep-ph/9209236
%From: GONDOLO@vand.physto.se
%Date: Tue, 15 Sep 1992 16:58 +0200

%
\input phyzzx
\catcode`\@=11
\def\andRefnum#1{\REFNUM #1}
\def\andRef#1{\andRefnum #1\REFWRITE }

\def\refitem#1{\r@fitem {[#1]} }
\def\figend@{\rel@x {\number\figurecount}}
\def\Fignum#1{\FIGNUM #1\figend@ }
\def\Fig#1{\Fignum #1\FIGWRITE }
\def\figitem#1{\r@fitem{Fig.~#1.}}
\catcode`\@=12

\def\|{\ifmmode\Vert\else \char`\|\fi}
\def\frac#1#2{{#1 \over #2}}
\def\simge{\rlap{\raise 2pt \hbox{$>$}}{\lower 2pt \hbox{$\sim$}}}
\def\simle{\rlap{\raise 2pt \hbox{$<$}}{\lower 2pt \hbox{$\sim$}}}

\def\a{ {\rm a} } \def\bel{ {\rm below} } \def\c{ {\rm c} } \def\CC{ {\rm CC} }
\def\cont{ {\rm cont} } \def\cut{ {\rm cut} } \def\d{ {\rm d} } \def\e{ {\rm e}
} \def\EAS{ {\rm EAS} } \def\eff{ {\rm eff} } \def\equal{ {\rm eq} } \def\F{
{\rm F} } \def\N{ {\rm N} } \def\NC{ {\rm NC} } \def\nucl{ {\rm nucl} } \def\p{
{\rm p} } \def\rock{ {\rm rock} } \def\th{ {\rm th} } \def\W{ {\rm W} } \def\Z{
{\rm Z} }

 \def\persec{ {\rm s}^{-1} } \def\yr{ {\rm yr} }

\def\eV{ {\rm eV} }  \def\TeV{ {\rm TeV} } \def\perTeV{
{\rm TeV}^{-1} }

\def\cm{ {\rm cm} } \def\sqcm{ {\rm cm}^2 } \def\persqcm{ {\rm cm}^{-2} }
\def\percc{ {\rm cm}^{-3} }

 \def\persr{ {\rm sr}^{-1} } \def\persqcmssr{ \persqcm \,
\persec \, \persr }

\def\g{ {\rm g} } \def\kton{ {\rm kton} } \def\perktonyr{ \kton^{-1} \, {\rm
yr}^{-1} }

%frontpage
\nopubblock
\vbox to 1in {
\hbox to 6.5in {\hfill UCLA/91/TEP/31 (revised)}
\hbox to 6.5in {\hfill August 1992}
\vss }
\titlepage
\title{Cosmic Neutrinos from Unstable Relic Particles}
\author{Paolo Gondolo}
\address{Department of Radiation Sciences, Uppsala University, P.O.\ Box~535,
 75121 Uppsala, Sweden}
\author{Graciela Gelmini}
\address{Department of Physics, University of California, \break
 Los Angeles, CA 90024-1547, USA}
\author{Subir Sarkar}
\address{Department of Physics, University of Oxford, \break
 1 Keble Road, Oxford OX1 3NP, UK}
\vskip 1cm
\abstract
We derive constraints on the relic abundance of a generic particle of mass
$\sim~1-10^{14}$ TeV  which decays into neutrinos at cosmological epochs, using
data from the Fr\'ejus and IMB nucleon decay detectors and the Fly's Eye air
shower array. The lifetime of such unstable particles which may constitute
the dark matter today is bounded to be greater than $\sim~10^{14}-10^{18}$ yr,
depending on the mass. For lifetimes shorter than the age of the universe,
neutrino energy losses due to scattering and the expansion redshift become
important and set limits to the ability of neutrino observatories to probe the
early universe.
\endpage

\chapter{Introduction}

Upper limits on the flux of high energy cosmic neutrinos obtained from nucleon
decay experiments and cosmic ray observatories constrain the relic cosmological
abundance of heavy unstable particles which decay into neutrinos. Given the
energy spectrum of the decay neutrinos and the decay branching ratio, upper
bounds can be obtained on the primordial abundance of the particle as a
function of its lifetime. Earlier attempts to set such bounds [\Ref\Frampton{
P.H.~Frampton and S.L.~Glashow, Phys.\ Rev.\ Lett.\ 44 (1980)
1481},\Ref\EllisGaisser{ J.~Ellis, T.~Gaisser and G.~Steigman, Nucl.\ Phys.\
B177 (1981) 427}] were made before any experimental data were available. We
present here the bounds imposed by the non-observation of extraterrestrial high
energy neutrinos in the Fr\'ejus and IMB nucleon decay detectors and the Fly's
Eye air shower array. We improve on previous work by taking into account the
experimental energy thresholds, the neutrino opacity of the early universe,
neutrino absorption in the Earth and the appropriate neutrino interaction cross
sections at high energies.

In section~2, we discuss the cosmological absorption of high energy neutrinos,
and in section~3 calculate the spectrum of neutrinos generated by heavy
particle decay. A discussion of the expected signals is given in section~4 and
the constraints provided by present observations are presented in section~5.
Our conclusions follow in section~6.

\chapter{Cosmological neutrino absorption}

High energy neutrinos can be absorbed in interactions with the relic thermal
neutrino background and with nucleons in the early universe. The dominant
processes are the annihilation of a high energy neutrino (or antineutrino) with
a background antineutrino (or neutrino) and its inelastic scattering off a
nucleon. We obtain below an analytical formula for the absorption redshift
$z_\a (E_\e)$ at which the neutrino opacity of the universe $s_\nu$ is unity
for a neutrino emitted with energy $E_\e$; less than a fraction $1/\e$ of the
neutrinos emitted at redshifts larger than $z_\a (E_\e)$ propagate to the
present epoch.

Consider a neutrino emitted with energy $E_\e$ at time $t_\e$ corresponding to
redshift $z_\e$. The cosmological neutrino opacity $s_\nu(t_\e, E_\e)$ is the
mean number of scatterings undergone by the neutrino in which it could have
been absorbed, given by
$$
s_\nu (t_\e, E_\e) = \int_{t_\e}^{\,t_0} {\d t \over \tau_\nu (t, E_\nu)}\,,
\eqn \eqi
$$
where $t_0 \sim 0.65 \times 10^{10} \,\yr\, ({\mit \Omega}_0 h^2)^{-1/2}$ is
the present age of the universe and $\tau_\nu (t, E_\nu)$ is the mean free time
between collisions at time $t$ (and redshift $z$) for a neutrino of energy
$E_\nu = E_\e (1 + z)/(1 + z_\e)$.  Here ${\mit \Omega}_0$ is the present mass
density of the universe in units of the critical density $\rho_{\c} \simeq 1.9
\times 10^{-29} h^2$ g $\percc$, where $h$ is the Hubble constant in units of
100 km $\persec$ Mpc$^{-1}$ ($0.4 \,\simle\, h \,\simle\, 1$). Taking into
account the two absorption processes mentioned above:
$$
{ 1 \over \tau_\nu(t,E_\nu) } =
{ 1 \over \tau_{\nu\bar\nu} } +
{ 1 \over \tau_{\nu \N} } \, ,
\eqn \eqii
$$
where the first term refers to neutrino-antineutrino annihilation and the
second to neutrino-nucleon scattering.

To obtain $\tau_{\nu\bar\nu}$, we must average over the thermal energy
distribution of the relic background (anti)neutrinos. Consider a decay neutrino
and a background antineutrino; the same formulae apply to a decay antineutrino
and a background neutrino. Indicating by $\theta_{\nu\bar\nu}$ the angle
between the two colliding particles in the cosmic frame, we have
$$
{ 1 \over \tau_{\nu\bar\nu} } =
\big\langle ( 1 - \cos \theta_{\nu\bar\nu} ) \sigma_{\nu\bar\nu}
\big\rangle \, n_{\bar\nu} \, ,
\eqn \eqiii
$$
where $(1 - \cos\theta_{\nu\bar\nu})$ is the $\nu\bar\nu$ relative velocity (in
units of $c$), $\sigma_{\nu\bar\nu}$ is the total $\nu\bar\nu$ annihilation
cross section and  $n_{\bar\nu}$ is the background antineutrino number density
at temperature $T_{\bar\nu}$, given by
$$
n_{\bar\nu} = {3 \over 4} {\zeta(3) \over \pi^2} T^3_{\bar\nu} \, .
\eqn \eqiv
$$
The angular brackets indicate an average over the antineutrino energy
distribution,
$$
f_{\bar\nu} (E_{\bar\nu}) =
{ 1 \over 2 \pi^2 }{ E_{\bar\nu}^2 \over \e^{E_{\bar\nu}/T_{\bar\nu}} + 1 }\, .
\eqn \eqivb
$$

We consider only annihilations into charged fermion pairs, $\nu\bar\nu \to f
\bar f$. In this case, $\sigma_{\nu\bar\nu} = \sum_f \sigma_{\nu\bar\nu \to f
\bar f}$, with the sum running over quarks and charged leptons. It is a good
approximation for our purposes to consider massless fermions to estimate the
annihilation cross section and simply add a new fermion channel whenever $E_\nu
T_{\bar\nu} > m_f^2$. In this case
$$
\sigma_{\nu\bar\nu} = { G_\F^2 s \over 4 \pi } \left[ N^\NC_\eff P_\Z(s) +
N^\CC_\eff A_\W(s) \right] \, ,
\eqn \eqv
$$
with $s = 2 E_\nu E_{\bar\nu} (1 - \cos \theta_{\nu\bar\nu})$. The first term
in the square brackets is due to neutral currents. The Z boson pole factor is
defined as
$$
P_\Z (s) =
{ M_\Z^4 \over ( s - M_\Z^2 )^2 + M_\Z^2 {\mit \Gamma}_\Z^2 } \ .
\eqn \eqvi
$$
The second term takes into account the charge current contribution to the
processes  $\nu_e \bar\nu_e \to e^+ e^- $, $\nu_\mu \bar\nu_\mu \to \mu^+
\mu^-$, $\nu_\tau \bar\nu_\tau \to \tau^+ \tau^-$:
$$
A_\W(s) = { M_\W^6 \over 2 s^3 \sin^2\theta_\W} \left[ 1 - { M_\W^2 \over
s + M_\W^2 } + (1-2a) {s \over M_\W^2} + {a s^2 \over M_\W^4} + 2(a-1)
\ln\left(
1 + {s\over M_\W^2}\right) \right] ,
\eqn \eqviz
$$
with
$$
a = \left( \coeff{1}{2} - \sin^2\theta_\W \right) { M_\Z^2 (s - M_\Z^2) \over
(s - M_\Z^2)^2+{\mit\Gamma}_\Z^2M_\Z^2 } \, .
\eqn \eqvizz
$$
The coefficients $N_\eff$ are the effective numbers of annihilation channels.
For neutral currents, this is calculated as
$$
N^\NC_\eff =
\sum_f \theta\!\left(E_{\nu} T_{\bar\nu} - m_f^2 \right)
\coeff{2}{3} n_f
\left( 1 - 8 t_{3f} q_f \sin^2 \theta_\W +
       8 q_f^2 \sin^4 \theta_\W \right) \, ,
\eqn \eqvii
$$
while for charged currents, the coefficient
$$
N^\CC_\eff = \theta\!\left(E_{\nu} T_{\bar\nu} - m_l^2 \right) \coeff{16}{3}
\sin^2\theta_\W
\eqn \eqviiz
$$
is non-zero only if the charged lepton $l$ is in the same family as the
annihilating neutrino. Above, $n_f$ is the number of colours (1 for leptons, 3
for quarks), and $t_{3f}$ and $q_f$ are the third component of the weak isospin
and the electric charge of the fermion in units of the positron charge
respectively. We take the electroweak mixing angle to be given by $\sin^2
\theta_\W = 0.23$. Inserting eq.~\eqv\ into eq.~\eqiii, we obtain
$$
{ 1 \over \tau_{\nu\bar\nu} } = { G_\F^2 \over 4 \pi } \bigg\langle ( 1 - \cos
\theta_{\nu\bar\nu} ) \left[ N^\NC_\eff  s P_\Z(s) + N^\CC_\eff s A_\W(s)
\right] \bigg\rangle \, n_{\bar\nu} \, ,
\eqn \eqviii
$$
where $s =  2 E_\nu E_{\bar\nu} ( 1 - \cos \theta_{\nu\bar\nu} )$ is
understood.

The thermal average has a peak at $E_\nu T_{\bar\nu} \simeq M_\Z^2 / 4$,
corresponding to the Z pole. All neutrinos emitted with energy  $E_\e
\,\simge\, M_\Z^2 / 4 T_{\bar\nu} = 1.26 \times 10^{13} \,\TeV / (1+z_\e)$ are
absorbed. For $ E_\e \,\simle\, M_\W^2 / 4 T_{\bar\nu}$, the  factors $P_\Z
(s)$ and $A_\W(s)$  in eq.~\eqv\ can be set equal to unity and $
\tau_{\nu\bar\nu} $ is easily evaluated as
$$
{ 1 \over \tau_{\nu\bar\nu} } =
\sigma_0 N_\eff E_\nu \rho_{\bar\nu} =
\rho_{{\bar\nu}_0} \sigma_0 N_\eff { (1+z)^5 \over 1+z_\e} E_\e \, .
\eqn \eqix
$$
Here, $N_\eff = N^\NC_\eff + N^\CC_\eff$ is the effective total number of
annihilation channels (which varies between 0.33 and 10.1), $\sigma_0$ is
defined as
$$
\sigma_0 \equiv { 2 \over 3 \pi } G_\F^2 =
1.12 \times 10^{-32} \,\sqcm \,\TeV^{-2} ,
\eqn \eqx
$$
and
$$
\rho_{{\bar\nu}_0} = 5.96 \times 10^{-14} \,\TeV \,\percc
\eqn \eqxi
$$
is the present antineutrino energy density (per species), taking the present
photon temperature to be 2.74 K [\Ref\COBE{ J.C.~Mather {\it et~al},
Astrophys.\ J.\ Lett.\ 354 (1990) L37}].  Thus, the annihilation mean free
time is
$$
\tau_{\nu\bar\nu} = 1.58 \times 10^{27} \,\yr \ N_\eff^{-1}  (1+z)^{-5}
(1+z_\e) \left(\frac{E_\e}{\TeV}\right)^{-1} \! .
\eqn \eqxii
$$

Next we consider neutrino-nucleon scattering. The thermal motion of the
non-relativis\-tic nucleons can be neglected, and we have
$$
{1 \over \tau_{\nu \N} } =
n_\N \sigma_{\nu \N} =
n_{\N_0} {\sigma_{\nu \N} \over E_\nu} { (1+z)^4 \over 1+z_\e } E_\e \, ,
\eqn \eqxiii
$$
where $n_\N$ is the nucleon number density at redshift $z$ and
$\sigma_{\nu \N}$ is the neutrino nucleon scattering cross section at
neutrino energy $E_\nu$. The present nucleon mean density is in the
range $n_{\N_0} \sim (0.25 - 1.5) \times 10^{-7} \,\percc$ according to
Big Bang nucleosynthesis calculations; this reflects the observational
uncertainty in primodial $^4{He}$ mass fraction, which is taken to be
$\sim 0.21-0.24$ [\Ref\Smith{ M.S.~Smith, L.H.~Kawano and R.A.~Malaney,
Caltech preprint OAP-716 (1992)}].

For $E_\nu \,\simle\, 1 \,\TeV$, the ratio $\sigma_{\nu \N} /E_{\nu}$ is
constant and equal to $0.67 \times 10^{-35} \,\sqcm$ $\,\perTeV$ for neutrinos
and to $0.34\times 10^ {-35} \,\sqcm \,\perTeV$ for antineutrinos
[\Ref\revpartprop{ Review of Particle Properties, Phys.\ Lett.\ B239 (1990) 1
[section\ III.76]}-\Ref\GaisserStanev{ T.K.~Gaisser and T.~Stanev, Phys.\
Rev.\ D30 (1984) 985}\andRef\McKayRalston{ D.W.~McKay and J.P.~Ralston, Phys.\
Lett.\ B167 (1986) 103}\andRef\QuiggRenoWalker{ C.~Quigg, M.H.~Reno and
T.P.~Walker, Phys.\ Rev.\ Lett.\ 57 (1986) 774}\andRef\GaisserGrillo{
T.K.~Gaisser and A.F.~Grillo, Phys.\ Rev.\ D36 (1987) 2752}\andRef\RenoQuigg{
M.H.~Reno and C.~Quigg, Phys.\ Rev.\ D37 (1988) 657}\andRef\Butkevich{
A.V.~Butkevich, A.B.~Kaidalov, P.I.~Krastev, A.V.~Leonov-Vendrovski and
I.M.~Zhele\-znykh, Z.\ Phys.\ C39 (1988) 241}]. The neutrino scattering mean
free time at these energies is
$$
\tau_{\nu \N} \simeq 10^{24} \,\yr \, (1+z)^{-4} (1+z_\e)
\left(\frac{E_\e}{\TeV}\right)^{-1} \! .
\eqn \eqxiv
$$
Comparing with the annihilation mean free time \eqxii, we see that inelastic
scattering upon nucleons dominates only at redshifts $1+z \,\simle\, 10^3
N_\eff^{-1}$. However, now $\tau_ {\nu} \simeq \tau _{\nu \N} \,\simge\,
10^{15} \,\yr \gg t_0$, i.e.\ the universe has already become transparent to
neutrinos. At higher neutrino energies, $\sigma_ {\nu \N}/E_{\nu}$ decreases
and neutrino-nucleon scattering is even less important, becoming negligible at
all redshifts for $E_\nu \,\simge\, 10^6 \,\TeV$. Thus inclusion of $\nu \N$
scattering affects $z_\a (E_\e)$ only marginally. For simplicity of
presentation, we do not therefore write it explicitly in the formulae below,
although we have included it in the numerical calculations.

The last ingredient necessary to compute the neutrino opacity is the
relationship between the age of the universe and the redshift:
$$
t = \cases{ t_0 (1+z)^{-3/2} , & for $z<z_\equal$, \cr
            t_0 (1+z_\equal)^{1/2} (1+z)^{-2} , & for $z>z_\equal$,
\cr}
\eqn \eqxv
$$
where $1+z_\equal = 2.25 \times 10^4 {\mit \Omega}_0 h^2$ is the redshift at
which the energy density of matter (with present density parameter ${\mit
\Omega}_0$) begins to dominate over that of radiation.

The absorption redshift obtained from integration of eq.~\eqi, using
eqs.~\eqix, \eqxiii\
and \eqxv, is shown as the diagonal full line in the $ E_\e -t_\e$
(or $E_\e - z_\e$) plane in figure~\Fig\fone{ The absorption redshift $z_\a$
(line 1) for cosmic neutrinos as a function of the neutrino energy at emission
$E_\e$ taking ${\mit \Omega}_0 h^2 = 1$. The other lines indicate: (2) the
boundary between the regions where absorption due to annihilation and
scattering dominate; (3) the present epoch; (4) the Z boson pole; (5) the epoch
of matter-radiation equality; (6) the epoch of light neutrino decoupling. }
taking ${\mit \Omega}_0 h^2 = 1$. The dot-dashed line separates the two regions
where annihilation and scattering absorption dominate. The region of interest
extends from the present epoch ($z = 0$), through the epoch of matter-radiation
equality ($z \simeq 2 \times 10^4$), up to the epoch of light neutrino
decoupling ($z \simeq 10^{10}$), which are all indicated. The location of the Z
boson pole is also shown as a diagonal dashed line.

Approximate expressions for the absorption redshift $z_\a(E_\e)$ can be
obtained for $ 1 \ll z_\e < z_\equal$ and $z_\e \gg z_\equal$. In these cases,
the result of the integration simplifies to
$$
s_\nu = \cases{
3.5 \times 10^{-17} ({\mit \Omega}_0 h^2)^{-1/2}
(1+z_\e)^{5/2}
\left( { E_\e / \TeV } \right) ,
& for $ 1 \ll z_\e < z_\equal $ , \cr
0.81 \times 10^{-14}
(1+z_\e)^2
\left( { E_\e / \TeV } \right) ,
& for $ z_\e \gg z_\equal $ . \cr}
\eqn \eqxvi
$$
The absorption redshift $z_\a(E_\e)$ is then obtained by setting $s_\nu = 1$:
$$
1+z_\a(E_\e) = \cases{
         3.8 \times 10^6 ({\mit \Omega}_0 h^2)^{1/5}
         \left( { E_\e / \TeV } \right)^{-2/5} \!\! ,
& $ E_\e \,\simge\, 5.2 \times 10^5 \,\TeV
({\mit \Omega}_0 h^2)^{-2} \! , $ \cr
         1.1 \times 10^7
         \left( { E_\e / \TeV } \right)^{-1/2} \!\! ,
& $ E_\e \,\simle\, 5.2 \times 10^5 \,\TeV
({\mit \Omega}_0 h^2)^{-2} \! .$\cr}
\eqn \eqxvii
$$

\chapter{Neutrino spectrum}

We now determine the present energy spectrum of neutrinos originating from the
decay of an unstable heavy particle $x$ with decay lifetime $\tau_x$. The
number of neutrinos of type $\nu_i$ ($\nu_i = \nu_\e, \bar\nu_\e, \nu_\mu,
\bar\nu_\mu, \dots $) produced at time $t$, per unit comoving volume and unit
time, is
$$
\gamma_\e(t) = { B_{\nu_i} Y_x(t) \over \tau_x }
= { B_{\nu_i} Y_{x_\p} \over \tau_x}
\,\exp\!\left( - { t \over \tau_x } \right)
= { B_{\nu_i} Y_{x_0} \over \tau_x}
\,\exp\!\left( - { t - t_0 \over \tau_x } \right) ,
\eqn \eqxviii
$$
where $B_{\nu_i}$ is the number of neutrinos of type $\nu_i$ produced per
decaying $x$ particle, $Y_x(t) \equiv n_x(t)/n_\gamma(t)$ is the $x$ particle
number density in ratio to the thermal photon density $n_\gamma(t)\ ( = 412.7
(1+z)^3 \,\percc )$, $Y_{x_\p}$ is its primordial value~\foot{This is
conveniently measured at the earliest epoch following which the comoving photon
number is conserved, say at $T \sim 0.01 m_\e$, corresponding to $t \sim
10^{-3} \ \yr$; this is negligible compared to all other time-scales relevant
here.} and $Y_{x_0}$ is its value today.

The number of neutrinos absorbed in the same volume, $\gamma_\a(t)$, is
proportional to the comoving density of decay-generated neutrinos $Y_{\nu_i}(t)
= n_{\nu_i}(t)/\allowbreak n_\gamma(t)$ and is given by
$$
\gamma_\a(t) = { Y_{\nu_i}(t) \over \tau_{\nu_i}(t,E_\e) } \, ,
\eqn \eqxix
$$
where $\tau_{\nu_i}(t,E_\e)$ is the neutrino absorption mean free time (see
section~2).

The evolution of the comoving neutrino density $Y_{\nu_i}(t)$ is governed by
$$
\eqalign{
{ \d Y_{\nu_i}(t) \over \d t} &=
\gamma_\e(t) - \gamma_\a(t) = \cr &=
{ B_{\nu_i} Y_{x_\p} \over \tau_x }
\,\exp\! \left( -{t \over \tau_x} \right) -
{ Y_{\nu_i}(t) \over \tau_{\nu_i}(t,E_\e) } \, . \cr}
\eqn \eqxx
$$
with the following solution at the present epoch $t_0$:
$$
Y_{\nu_{i_0}} = B_{\nu_i} Y_{x_\p}
\, \int _0 ^{t_0}
\exp \left[ - {t_\e \over \tau_x} - s_{\nu_i}(t_\e,E_\e) \right]
\, {\d t_\e \over \tau_x} \, .
\eqn \eqxxi
$$

Now differentiating with respect to the present neutrino energy $E_{\nu_{i_0}}
= E_\e (1 + z_\e)^{-1}$, one obtains the present neutrino flux
$$
E_{\nu_{i_0}} { \d \phi_{\nu_{i_0}} \over \d E_{\nu_{i_0}} } =
\phi_{\gamma_0} B_{\nu_i} Y_{x_\p} \, \kappa \ {t_\e \over \tau_x}
\exp \left[ -{t_\e \over \tau_x} - s_{\nu_i}(t_\e,E_\e) \right]
\theta (E_\e - E_{\nu_{i_0}}) ,
\eqn \eqxxii
$$
where
$$
t_\e = t_0 \left( {E_{\nu_{i_0}} \over E_\e} \right)^\kappa , \qquad
\kappa = \cases{ 2,
     & for $E_{\nu_{i_0}} < E_\e (1+z_\equal)^{-1}$, \cr
                 \coeff{3}{2},
     & for $E_{\nu_{i_0}} > E_\e(1+z_\equal)^{-1}$, \cr}
\eqn \eqxxiii
$$
and $\phi_{\gamma_0} = n_{\gamma_0} / 4 \pi = 0.98 \times 10^{12}
\,\persqcmssr$ is the present background photon flux per unit solid angle. The
decay neutrino flux is shown in figure~\Fig\ftwo{The present energy spectrum of
decay generated neutrinos for $\tau_x = 10^{-5} t_0$ and $3 E_\e =
10^5 \,\TeV$ (line 1), $10^7 \,\TeV$ (line 2) and
$10^9 \,\TeV$ (line 3).
The full lines show the effects of
cosmological neutrino absorption.} for $\tau_x = 10^{-5} t_0$ and $3 E_\e =
10^5
\,\TeV$, $10^7 \,\TeV$, $10^9 \,\TeV$. Notice that the present neutrino energy
$E_\nu$ is redshifted from $E_e$.
The dotted lines indicate what the flux
would have been without cosmological neutrino absorption. These three curves
are simple translations of each other, since the differential flux~\eqxxii\
with $s_{\nu_i}=0$ depends only on the ratio $E_{\nu_{i_0}}/E_\e$.

Approximating the effect of the cosmological neutrino absorption with
$\e^{-s_{\nu_i}} \simeq \theta(t_\e - t_\a)$, where $t_\a < t_0$ corresponds to
the absorption redshift $z_\a(E_\e)$ at which the neutrino opacity is unity,
eq.~\eqxxii\ can be easily integrated to obtain the total neutrino flux today,
$$
\phi_{\nu_{i_0}} \simeq
\phi_{\gamma_0} B_{\nu_i} Y_{x_\p}
( \e^{-t_\a/\tau_x} - \e^{-t_0/\tau_x} ) .
\eqn \eqxiv
$$
For $\tau_x \ll t_0 - t_\a$, this reduces to
$$
\phi_{\nu_{i_0}} \simeq
\phi_{\gamma_0} B_{\nu_i} Y_{x_\p} \e^{-t_\a/\tau_x} ,
\eqn \eqxxva
$$
while for $\tau_x \gg t_0 - t_\a$, it becomes
$$
\phi_{\nu_{i_0}} \simeq
\phi_{\gamma_0} B_{\nu_i} Y_{x_0} {t_0-t_\a \over \tau_x} \, .
\eqn \eqxxvb
$$
The neutrino flux is exponentially suppressed for $\tau_x \ll t_\a$ and reaches
a maximum of $\phi_{\nu_{i_0}} \simeq B_{\nu_i} Y_{x_\p} \times 10^{12}
\,\persqcmssr$ for $\tau_x \simeq t_0 - t_\a$. This flux is potentially
enormous compared with present bounds on the diffuse extragalactic high energy
neutrino flux ($\sim~10^{-6} \,\persqcmssr$ for $E_{\nu} \,\simge\, 1 \,\TeV$
and $\sim~10^{-16} \,\persqcmssr$ for $E_{\nu} \,\simge\, 10^7 \,\TeV$) which
may be inferred from data obtained with underground detectors and cosmic ray
observatories (cf.\ section~4). Hence very restrictive bounds may be obtained
on the abundance of the hypothetical decaying particle as demonstrated below.

\chapter{Expected signals}

At present, the best means to detect a diffuse background of high
energy neutrinos is through the production of an energetic charged
lepton in the collision of such a neutrino with a nucleon. We consider
three possible types of signal according to where the interaction
occurs. An event is called {\sl contained} when the interaction occurs
inside an underground detector, such as Fr\'ejus, IMB and Kamiokande. A
flux of {\sl through-going muons} is registered when interactions occur
in the material surrounding the detector (rock in the underground
experiments and water in the forthcoming DUMAND, GRANDE and Lake Baikal
experiments). Finally, if the interaction occurs in the atmosphere an
{\sl extensive air shower (EAS)} is generated, which can be detected by
cosmic ray observatories such as Fly's Eye and CASA.

We consider the following experimental constraints on the {\it total} neutrino
flux:

(1) the rate of contained events in the Fr\'ejus detector, with electron and/or
muon energies greater than 3 GeV, does not exceed 17.7 $\perktonyr$
[\Ref\Frejus{ Ch.~Berger {\it et~al}, Phys.\ Lett.\ B227 (1989) 489}];~\foot{We
consider the 11 electron and 14 muon charged current events over 3 GeV observed
in 1.56 kton yr (see fig. 3 of ref.~[\Frejus]), and apply the quoted
identification efficiencies of 85\% and 95\% respectively.}

(2) the rate of contained events with energies between 100 MeV and 2.5 GeV in
the IMB-3 detector is limited by 111.5 $\perktonyr$ [\Ref\IMBc{ D.~Casper {\it
et~al}, Phys.\ Rev.\ Lett.\ 66 (1991) 2561}].~\foot{From the total number of
422 contained events in 3.4 kton yr, we exclude the 43 events below 100 MeV
(see fig. 2 of ref.~[\IMBc]) where the track reconstruction and identification
efficiencies are low. For comparison the IMB-1 detector recorded 401 contained
events in 3.77 kton yr [\Ref\IMBone{ R.M. Bionta {\it et~al}, Phys.\ Rev.\ D38
 (1988) 768}].}

\noindent
In fact, the observed contained events are well accounted for by the expected
neutrino flux from cosmic ray interactions in the atmosphere [\Ref\Barr{
T.K.~Gaisser, T.~Stanev and G.~Barr, Phys.\ Rev.\ D38 (1988) 85; \nextline
G.~Barr, T.K.~Gaisser and T.~Stanev, Phys.\ Rev.\ D39
(1989) 3532; \nextline E.V.~Bugaev and V.A.~Naumov,
Phys. Lett. B232 (1989) 391; \nextline H.~Lee and
Y.~Koh, Nuovo Cimento B105 (1990) 883; \nextline M.~Honda {\it et~al},
Phys.\ Lett.\ B248 (1990) 193}], within the uncertainty of
$\sim 25\%$ in these computations. Hence the bound on contained events of
non-atmospheric origin can, in principle, be improved by up to a factor of
$\sim
10$ and the limits to be derived strengthened proportionally.

We also consider the following constraints on any {\it extraterrestrial}
neutrino flux:

(3) the flux of upward-going muons (from directions with zenith angle
larger than  98$^\circ$) with energy greater than 2 GeV registered by
the IMB-1 detector is less than $2.65 \times 10^{-13} \,\persqcm
\,\persec$ at the 90\% confidence level, after subtraction of the
expected atmospheric component [\Ref\IMBmu{ R.~Svoboda {\it et~al},
Astrophys.\ J.\ 315 (1987) 420}];\foot{The recent Kamiokande
upper limit of $4 \times 10^{-14} \,{\rm muons} \,\persqcm \,\persec
\,\persr$ for zenith angles larger than 150$^\circ$
[\Ref\Kamiokande{M.~Mori {\it et~al},  Phys.\ Lett.\ B278 (1992) 217}]
is slightly less stringent than the IMB limit we consider, which
corresponds to $3.7 \times 10^{-14} $ $ \,{\rm muons} \,\persqcm \,\persec
\,\persr$.}

(4) the Fly's Eye array has set upper limits on the rate of neutrino-induced
EAS's of $10^{-45}\,\persec \,\persr$, $3.8 \times 10^{-46}\,\persec \,\persr$,
 $10^{-46} \,\persec \,\persr$ and  $3.8 \times 10^{-47} \,\persec \,\persr$,
all at the 90\% confidence level, for neutrino energies higher than
$10^5\,\TeV$, $10^6\,\TeV$, $10^7\,\TeV$ and $10^8\,\TeV$ respectively
[\Ref\FlysEye{ R.M.~Baltrusaitis {\it et~al}, Phys.\ Rev.\ D31 (1985) 2192}].

For the isotropic neutrino flux~\eqxxii, the rate of contained events per unit
detector mass, $R_c$, the upward-going muon flux, $\phi_\mu$, and the rate of
EAS's per unit solid angle, $J$, can all be written in the form
$$
S = \sum_i \int \d E_{\nu_i}
{ \d \phi_{\nu_i} \over \d E_{\nu_i} } \,
P_i(E_{\nu_i}) \,
\Omega_i(E_{\nu_i}) \, ,
\eqn \eqxxvi
$$
where the signal $S$ is $R_c$, $\phi_\mu$ or $J$, and the sum is over neutrino
types ($\nu_i = \nu_\e$, $\bar\nu_\e$, $\nu_\mu$, $\bar\nu_\mu$, $\dots$). The
effective aperture $\Omega_i(E_{\nu_i})$, which takes account of neutrino
absorption by the Earth (if any), and the transfer functions $P_i(E_{\nu_i})$
depend on the experimental data set considered. (We have dropped the subscript
0 referring to the present neutrino energy.)

For a simplified model of the Earth with uniform density $\rho_\oplus = 5.5
\,\g \,\percc$ and radius $R_\oplus = 6.37 \times 10^8 \,\cm $, the effective
aperture is (neglecting the depth of the underground detector relative to
$R_\oplus$),
$$
\Omega_i(E_{\nu_i}) =
\int \d \Omega
\exp\! \big[ - 2 R_\oplus k_i(E_{\nu_i}) | \cos \vartheta | \big]
\, \theta(-\cos\vartheta) ,
\eqn \eqxxvii
$$
where the integral extends over the geometrical aperture of the detector,
$\vartheta$ is the zenith angle and $k_i(E_{\nu_i})$ is the neutrino absorption
coefficient in the Earth, given by
$$
k_i(E_{\nu_i}) =
{\rho_\oplus \over m_\N }
\sigma_{\nu_i \N} (E_{\nu_i}) ,
\eqn \eqxxviii
$$
with $m_\N$ the nucleon mass and $\sigma_{\nu_i \N} (E_{\nu_i})$ the total
neutrino-nucleon cross section. For the neutrino energies under consideration,
$\sigma_{\nu_i \N} (E_{\nu_i})$ includes only the charged current cross section
$ \sigma_{\nu_i \N}^{\CC} (E_{\nu_i})$, because the energy and momentum
fractions transferred to the nucleon in a neutral current process are
negligible at these energies.

The charged current cross section $\sigma_{\nu_i \N}^{\CC} (E_{\nu_i})$ is
well-known for $ E_{\nu_i} \,\simle\, 10 \,\TeV$:
$$
\sigma_{\nu_i \N}^{\CC} = 0.67 \times 10^{-35} \,\sqcm
\left( {E_{\nu_i} \over \TeV} \right) \, ,
\eqn \eqxxviib
$$
for a neutrino and
$$
\sigma_{\bar\nu_i \N}^{\CC} = 0.34 \times 10^{-35} \,\sqcm \left( {E_{\nu_i}
\over \TeV} \right) \, ,
\eqn \eqxxviib
$$
for an antineutrino [\revpartprop]. At higher energies the charged current
cross
section becomes more and more uncertain --- by as much as a factor of 10 at
$E_{\nu_i} \simeq 10^9 \,\TeV$ --- because of the poor knowledge of nucleon
structure functions at small arguments [\Butkevich]. For this reason, we have
not attempted a precise calculation of $\sigma_{\nu_i \N}^{\CC} (E_{\nu_i})$
from a set of theoretical structure functions. We have used the differential
charged current cross section up to $ E_{\nu_i} = 10^7 \,\TeV$ given in
ref.~[\QuiggRenoWalker]. At still higher energies, we have matched the
asymptotic form of the cross section in ref.~[\McKayRalston] to the results of
ref.~[\QuiggRenoWalker].

In figure~\Fig\fthree{The effective detector aperture, integrated below the
horizon, for neutrinos (solid line) and antineutrinos (dotted line),
demonstrating the opacity of the Earth at high energies. } we show the effect
of absorption in the Earth by plotting the effective aperture \eqxxvii\
integrated below the horizon:
$$
\Omega_i^\bel (E_{\nu_i}) =
{ 2 \pi \sigma_\oplus \over \sigma_{\nu_i \N }(E_{\nu_i}) }
\left[ 1 - \exp\!\left( - {\sigma_{\nu_i \N}(E_{\nu_i}) \over
\sigma_\oplus } \right) \right] ,
\eqn \eqxxix
$$
with $\sigma_\oplus = m_\N / 2 R_\oplus \rho_\oplus = 2.4 \times 10^{-34}
\,\sqcm $. This effective aperture differs very little from the one obtained in
ref.~[\Ref\Berezinski{ V.S.~Berezinskii, A.Z.~Gazizov, G.T.~Zatsepin and
I.L.~Rozental, Soviet J.\ Nucl.\ Phys.\ 43 (1986) 637}] using a more elaborate
model of the Earth. As we see from the figure, the Earth severely attenuates
the flux of neutrinos of energy exceeding $\sim 10^5$ TeV, becoming nearly
opaque at $\sim 10^{10}$ TeV.

Note that for $\sigma \,\simge\, 10^{-33} \,\sqcm$, i.e.\ at $E_{\nu_i}
\,\simge\, 10^3 \,\TeV $, the effective aperture from below the horizon is
inversely proportional to the neutrino-nucleon scattering cross section,
$$
\Omega_i^\bel (E_{\nu_i} ) \simeq
2 \pi { \sigma_\oplus \over  \sigma_{\nu_i \N}
(E_{\nu_i}) } \, .
\eqn \eqxxx
$$

The resonant reaction $\bar\nu_\e e^- \to \W^- \to $ ``anything'' severely
depletes the $\bar\nu_\e$ flux from below the horizon at energies around
$E_{\bar\nu_\e} \simeq 7 \times 10^3 \,\TeV$ [\Ref\Glashow{ S.L.~Glashow,
Phys.\ Rev.\ 118 (1960) 316}]. However since we do not assume any predominant
neutrino type in the decay neutrino flux, the $\bar\nu_\e$ flux from below
accounts for only one eighth of the total rate of contained events. It is
therefore a reasonable approximation, for our purposes, to neglect this
resonance.

We present now the transfer functions $P_i(E_{\nu_i})$ for the experimental
data sets under consideration. For contained events we have
$$
P_i^\cont(E_{\nu_i}) =
\theta(E_{\nu_i}-E_\th)
{ N_\nucl \over M}
\int_{E_\th}^{\min(E_\cut, E_{\nu_i})}
\d E_{l_i}
{ \d \sigma^{\CC}_{\nu_i \N} \over \d E_{l_i} } ,
\eqn \eqxxxi
$$
where $E_\th$ is the experimental energy threshold for the lepton energy
$E_{l_i}$, $E_\cut$ is an experimental cutoff (2.5 GeV for IMB and infinite for
Fr\'ejus), $M$ is the detector mass and $N_\nucl = 6.02 \times 10^{32}\,
(M/\kton)$ is the number of nucleons in the detector. If the charged current
cross section is written in units of $10^{-38} \,\sqcm$, $\sigma^{\CC}_{\nu_i
\N} (E_{\nu_i}) = \sigma_{i,38} \times 10^{-38} \,\sqcm$, then the transfer
function for contained events is
$$
P_i^\cont(E_{\nu_i}) \simeq 6.0 \times 10^{-6} \,\sqcm\,\kton^{-1}\
\theta(E_{\nu_i}-E_\th)
\int_{E_\th}^{\min(E_\cut, E_{\nu_i})}
 \d E_{l_i} { \d \sigma_{i,38}
\over \d E_{l_i} } .
\eqn \eqxxxii
$$
Both Fr\'ejus and IMB data sets include neutrinos coming from all solid angles
and their effective aperture computed from eq.~\eqxxvii\ varies from $4\pi$ to
$2\pi$ as the energy is increased; the neutrino flux is reduced at most by a
factor of 2 at the highest energies.

In IMB, and other water-\v{C}erenkov detectors, there is also the possibility
that the hadronic fragments produced in the neutrino-nucleus collision give a
detectable amount of \v{C}erenkov light. Their contribution  $P_{i,{\rm
hadr}}^\cont(E_{\nu_i})$ should then be added to eq.~\eqxxxii.  In the
appendix, we present an estimate of the contribution from such
`hadronic blasts' and show that this is
important only for very energetic neutrinos, $E_{\nu_i} \simge 10^7 \,\TeV$,
where, however, the signal from EAS's gives more stringent constraints.

The product $P^\cont_i(E_{\nu_i})\
\Omega^\cont_i(E_{\nu_i})$ thus obtained for the Fr\'ejus and IMB contained
events is shown in figure~\Fig\ffour{The product of the transfer function
$P_i(E_{\nu_i})$ and of the effective aperture $\Omega_i(E_{\nu_i})$ for the
three experimental data sets we consider: (a) contained events in IMB (curve 1)
and Fr\'ejus (curve 2) (b) IMB upward-going muons and (c) Fly's Eye EAS's
(four thresholds). The dotted lines corresponds to antineutrinos.  }(a) for
neutrinos (solid lines) and antineutrinos (dotted lines) as function of the
neutrino (or antineutrino) energy $E_{\nu_i}$. The units are chosen such that
the vertical axis directly gives the number of events per kiloton-year
corresponding to a unit neutrino flux of 1 $\persqcmssr$. For $E_{\nu_i}
\,\simge\, 10^6 \,\TeV$ we calculate $P^\cont_i (E_{\nu_i})\,\Omega^{\cont}_i
(E_{\nu_i}) \simeq 3.8 \times 10^{-5}$ cm$^2$  sr kton$^{-1}$ $\sigma_{i,38}$
for the Fr\'ejus detector. The IMB curve (curve 1) above $10^7 \,\TeV$ is due
to `hadronic blasts' as discussed in the appendix.

The transfer function for the flux of up-going muons is (see ref.
[\Ref\Gaisser{
T.K.~Gaisser, Cosmic Rays and Particle Physics (Cambridge Univ.\
Press, Cambridge, 1990)}])
$$
P^\mu_i(E_{\nu_i}) = \theta(E_{\nu_i}-E_\th) \int_{E_\th} ^\infty \d E'_\mu
\int_{E'_\mu} ^{E_{\nu_i}} \d E_\mu \int _0 ^\infty \d X\, g\,(X,E'_\mu,E_\mu)\
{ \d \sigma_{\nu_i \N} \over \d E_\mu } ,
\eqn \eqxxxiii
$$
for $\nu_i=\nu_\mu,\bar\nu_\mu$, and  $P^\mu_i(E_{\nu_i}) = 0$ for the other
neutrino types. Here $E_\mu$ and $E'_\mu$ are the muon energies at production
and at the detector respectively, and $X = l \, \rho_{\rock}/m_\N$ is the
column density of rock, i.e.\ the number of nucleons per unit area encountered
by a muon travelling a length $l$ in rock. Notice that $P^\mu_i$ is
adimensional.
The probability that a muon with
initial energy $E_\mu$ has an energy between $E'_\mu$ and $E'_\mu + \d E'_\mu$
after traversing an amount $x$ of rock is denoted by $g\,(X,E'_\mu,E_\mu)\,\d
E'_\mu$. We assume a uniform rock density of $\rho_\rock = 2.6 \ \g \,\percc$
in the region surrounding the detector. Following ref.~[\GaisserStanev], we
make
the approximation that the final muon energy $E'_\mu$ coincides with its mean
value (with no dispersion):
$$
\overline{E'_\mu} =
( E_\mu + \epsilon ) \e^{-\gamma X} - \epsilon ,
\eqn \eqxxxiv
$$
with $\epsilon \simeq 0.51 \,\TeV$ and $\gamma^{-1} = 1.54 \times 10^{29}
\,\persqcm$. The integral over $E'_\mu$ in eq.~\eqxxxiii\ can then be
performed, and we obtain
$$
P^\mu_i(E_{\nu_i}) = \theta(E_{\nu_i}-E_\th) \int_{E_\th} ^{E_{\nu_i}} \d E_\mu
X_\th (E_\mu) { \d \sigma_{\nu_i \N} \over \d E_\mu } ,
\eqn \eqxxxv
$$
for $\nu_i = \nu_\mu, \bar\nu_\mu $. Here
$$
X_\th (E_\mu) = \gamma^{-1} \ln \left( { 1 + E_\mu / \epsilon \over 1 + E_\th /
\epsilon } \right)
\eqn \eqxxxvi
$$
is the column density traversed by muons produced with energy $E_\mu$ which
reach the detector with threshold energy $E_{\rm th}$. A plot of
$P^{\mu}_i(E_{\nu_i})$ times $ \Omega^{\mu}_i(E_{\nu_i})$, obtained by
integrating eq.~\eqxxvii\ over zenith angles larger than 98$^\circ$, is shown
in figure~\ffour(b).
The decrease of
$P^{\mu} \Omega^{\mu}$
 for $E_{\nu_i} \,\simge\, 10^7 \,\TeV$
is due to absorption by the Earth.

The final signal we consider is the rate of EAS's per unit solid angle. Its
transfer function is
$$
P^\EAS_i(E_{\nu_i}) =
{1 \over \Omega_i^\EAS(E_{\nu_i})} \sigma_{\nu_i \N} (E_{\nu_i})
\theta(E_{\nu_i}-E_\th) ,
\eqn \eqxxxvii
$$
where the index $i$ stands for $\nu_e$ and $\nu_\tau$, which can
generate electromagnetic or hadronic showers in the atmosphere.  Muons
from charged current $\nu_\mu$-nucleon interactions do not trigger
electromagnetic showers, since their radiation length for bremstrahlung
in air ($10^5$ g/cm${}^2$) is much larger than the atmosphere thickness
(1030 g/cm${}^2$).  The product
$P^\EAS_i(E_{\nu_i})\,\Omega^\EAS_i(E_{\nu_i})$ is shown in
figure~\ffour(c) for the four experimental energy thresholds of the
Fly's Eye detector [\FlysEye].

We are now in a position to compare the expected signals to the experimental
limits.

\chapter{Present constraints}

We assume that the same numbers of neutrinos and antineutrinos of each type, $
B_{\nu_\e} = B_{\bar\nu_\e} = B_{\nu_\mu} = B_{\bar\nu_\mu} = \dots \equiv
B_\nu $, are produced in $x$ decays and that their production energy is always
$ E_\e = \coeff{1}{3} m_x $. Inserting the decay-generated neutrino flux,
eq.~\eqxxii, into eq.~\eqxxvi\ and using the appropriate transfer functions and
effective apertures described in section~3, we obtain the expected signals in
terms of the decay lifetime $\tau_x$ and the quantity $ B_\nu m_x Y_{x_\p} =
B_\nu m_x Y_{x_0} \e^{t_0/\tau_x}$ which is proportional to the primordial
energy density of the decaying particles. Notice that for $\tau_x \gg t_0$,
this quantity is $\sim 25.5 \,\eV\, (B_\nu \,{\mit \Omega}_{x_0} h^2) $, where
${\mit \Omega}_{x_0}$ is the present $x$ mass density in units of the critical
cosmological density.

We present the results in figure~\Fig\ffive{The bound on the primordial energy
density of the decaying particle multiplied by the branching ratio into
neutrinos, $B_\nu m_x Y_{x_\p}$, as a function of its lifetime $\tau_x$, for
various choices of its mass $m_x$. The shaded regions are excluded by the
present experimental data. The various lines refer to: IMB upward-going muons
(solid lines), Fr\'ejus and IMB contained events (dotted and short-dashed
lines respectively), Fly's Eye EAS's (long-dashed lines). Also indicated are
the upper bound on the total energy density ${\mit \Omega}_0 h^2 = 1 $
(short-dashed--dotted line) and the upper bound inferred from considerations
of primordial light element abundances (long-dashed--dotted line). }. The
shaded regions are excluded by the present experimental data. The solid lines
refer to the limit on upward-going muons from IMB, the dotted and short-dashed
lines to the Fr\'ejus and IMB contained events, respectively, and the
long-dashed lines to the Fly's Eye EAS's. These are essentially bounds on the
relic energy density of the decaying particle, taking $B_{\nu} = 1$; for $
B_{\nu} < 1$, these lines are to be proportionally shifted upward.  As noted
earlier, the experimental limits on contained events can, in principle, be
improved by a factor of $\sim 10$ if the signal due to atmospheric neutrinos
is accounted for; the corresponding bounds should then be scaled downwards by
the same factor. The short-dashed--dotted line corresponds to a present mass
density ${\mit \Omega}_0 h^2 = 1 $ either in $x$ particles (for $\tau_x
\,\simge\, t_0$) or in its decay products (for $\tau_x \,\simle\, t_0$); the
region ${\mit \Omega}_0 h^2 > 1$ is excluded by the observational lower limits
to the age and present expansion rate of the universe.  For comparison we also
show as a long-dashed--dotted line (in the lower left quadrant) the upper
bound on $m_x Y_{x_\p}$ for unstable particles decaying into
electromagnetically interacting particles. This is obtained by requiring that
the abundance of the primordially synthesised light elements $D,\ ^3{He},\
^4{He}$ and $^7{Li}$ not be excessively altered from their observationally
inferred values by the electromagnetic  cascades initiated by the decay
products [\Ref\Sarkar{ J.~Ellis, G.B.~Gelmini,  J.L.~Lopez, D.V.~Nanopoulos
and S.~Sarkar, Nucl.\ Phys.\ B373 (1992) 399}]. In fact this bound also
applies to unstable particles decaying into neutrinos, since the decay
neutrinos can initiate similar electromagnetic cascades through the process $
\nu \bar\nu \to e^+ e^- $, where the target (anti)neutrinos belong to the
thermal background. (This has also been considered in
ref.~[\Ref\Gratsias{J.~Gratsias, R.J.~Scherrer and D.N.~Spergel, Phys.\ Lett.\
B262 (1991) 298}]; however these authors do not calculate cascade generation
correctly and obtain an overly restrictive bound.)

Figures~\ffive (a-d) correspond to $m_x = 1,\ 10^5,\ 10^6$ and $10^{10}
\,\TeV$ respectively. As the $x$ mass increases, the bounds at $\tau_x
\,\simle\, t_0$ first shift to the left as the decay neutrinos become
more energetic and the signals go further above the experimental
thresholds. Then they proceed to move to the right because the neutrino
absorption redshift decreases with increasing neutrino energy, hence
the neutrinos produced by relatively short-lived particles do not
survive until the present. The mass at the turning point is given
approximately by solving  $m_x \simeq 3 E_\th [ 1 +
z_\a(\coeff{1}{3}m_x) ]$ with the help of eqs.~\eqxvii; its value is
$1\times 10^3 \,\TeV$ and $1\times 10^2 \,\TeV$ for contained events in
Fr\'ejus and IMB respectively, $8\times 10^2 \,\TeV$ for the IMB
upward-going muons, and $1\times 10^8\,\TeV$, $1\times 10^9 \,\TeV$,
$5\times 10^9 \,\TeV$ and $3\times 10^{10} \,\TeV$ for the four Fly's
Eye thresholds.

When $\tau_x \,\simge\, t_0$, the best bounds on the relic abundance of
the decaying particles come from the IMB limit on upward-going muons at
$ m_x \,\simle\, 5 \times 10^5 \,\TeV$ and the Fly's Eye limits on EASs
at $ m_x \,\simge\, 5 \times 10^5 \,\TeV$. We can invert the argument
and consider the interesting case $ {\mit \Omega}_{x_0} h^2 \simeq 1$,
i.e.\ when the $x$ particles are {\it assumed} to constitute the dark
matter today.\foot{In this case the actual spatial distribution of the
relic particles should be taken into account, e.g. their likely
concentration in the halo of our Galaxy. This would yield even stricter
constraints. Preliminary work has been reported in
ref.~[\Ref\Gond{P.~Gondolo, Uppsala preprint PT17-1992}] and a more
detailed study is in progress.} A corresponding lower bound on its
lifetime versus its mass can then be inferred and is plotted in
figure~\Fig\fmxtx{The lower bound on the $x$ particle lifetime versus
its mass for a present density ${\mit \Omega}_{x_0} h^2 = 1$ in the
relic particles, assuming unit branching ratio into neutrinos}. For
$m_x \,\simle\, 30 \,\TeV $, this bound gets stronger with increasing
$m_x$ as the mean energy of the decay generated neutrinos rises over
the IMB detection threshold.  For $m_x \,\simge\, 30 \,\TeV$, the lower
bound on $\tau_x$ is inversely proportional to $m_x$
(cf.\ eq.~\eqxxvb), apart from the jumps at $\simeq 10^5 \,\TeV$,
$\simeq 10^6 \,\TeV$, $\simeq 10^7 \,\TeV$ and $\simeq 10^8 \,\TeV$
corresponding to the Fly's Eye energy thresholds. No bound exists for
$m_x \,\simge\, 5\times10^{14} \,\TeV$, since the universe is opaque to
such high energy neutrinos at the present epoch.

\chapter{Conclusions}

We have considered constraints on the lifetime, the abundance and the mass of
unstable relic particles decaying into neutrinos at cosmological epochs, taking
into account that both the early universe and the Earth are opaque to very high
energy neutrinos. We have evaluated the signals expected from a diffuse
background of decay-generated neutrinos in underground nucleon decay
experiments and at cosmic ray observatories. Comparing these to the present
limits on the flux of non-atmospheric neutrinos, we find severe bounds on the
relic abundance of such heavy particles; in particular, such particles must be
very long-lived indeed in order to constitute the dark matter today. These
bounds are of relevance to massive metastable particles such as technicolour
baryons and `cryptons' (bound states in the hidden sector of
superstring-inspired models) as discussed elsewhere [\Sarkar].

\chapter{Acknowledgements}

We are grateful to the referee for suggesting that we consider the effects of
hadronic `blasts' in  water-\v{C}erenkov detectors.

This work was supported in part by the Department of Energy under the
contract DE-AT03-88ER-40384 Mod A006-Task C.

\bigskip
\centerline{\titlestyle Note Added}
\smallskip

Recently, we became aware of ref.~[\Ref\Berezinskii{V.~Berezinsky, Gran Sasso
preprint LNGS-91-02}], where the detection of neutrinos from cosmic
relic particles is also studied, in particular the effects due to
absorption in the early Universe. However this work assumes that the
decaying particles were thermally produced (with a calculable
abundance) in the early Universe, whereas we have presented results in
a general form, applicable to any relic particle. This, in fact, is
essential in order to consider particles with masses over a few hundred
TeV. We also believe that we have addressed experimental issues in more
detail.

\appendix

Here we present an estimate of the contribution of `hadronic blasts' to
the IMB transfer function for contained events. We calculate the visible
energy equivalent to the \v{C}erenkov light output from such a `blast' and then
compare it to the visible energy observable in the IMB detector.

Let $W$ be the energy transferred to the nucleus in the neutrino interaction.
This is also presumably the energy available to the hadronic shower generated
by the nuclear fragments. The visible energy $E_{\rm vis}$ is defined to be
the energy of a  fictitious initial electron generating an electromagnetic
shower with the same \v{C}erenkov light output [\IMBc]. The physics of
electromagnetic showers [\Ref\Rossi{Ref.~[\revpartprop] [section III.15]}]
then relates $E_{\rm vis}$  to the `detectable' track length $X_d$,
$$
E_{\rm vis} = E_c { X_d(W)  \over X_0 F(z)},
\eqn \ai
$$
where $E_c$ is the critical energy separating the domains where
ionization and radiation energy losses dominate and is approximately
given by
$$
E_c \simeq { 800 \,{\rm MeV} \over Z+1.2} =  71 \, {\rm MeV}.
\eqn \aii
$$
Here $X_0 (= 36.08$ g/cm${}^2$ [\Ref\radlength{ Ref.~[\revpartprop]
[section III.5]}]) is the electron radiation length in water and $F(z)$ is
approximately
$$
F(z) \simeq e^z \left( 1 + z \ln { z \over 1.526 } \right) ,
\eqn \aiii
$$
with
$$
z \simeq 4.58 { Z \over A} { E_d \over E_c} \simeq { E_d
\over  28 \,{\rm MeV}}
\eqn \aiv
$$
(the numerical values refer to water). In the IMB detector,
$E_d = 1.52 \, m_e + 30 \, {\rm MeV} + 140 \, {\rm MeV} = 170 \, {\rm
MeV}$ for an electron [\IMBone], hence  eq.~\ai\ reads
$$
E_{\rm vis} = 4.8 \times 10^{-4} \, {\rm MeV} \, X_d(W)
\eqn \av
$$
with $X_d(W)$ in g/cm${}^2$.

It remains now to estimate $X_d(W)$, the path length of charged
particles in the hadronic shower with energy above the detection
threshold. The mean number of charged particles $ n_{\rm ch}(W) $ as a
function of the available energy $W$ has been studied in deep inelastic
scattering and incorporated into the Lund Monte Carlo program
[\Ref\Ingelman{M.~Bengtsson, T.~Sj\"ostrand and G.~Ingelman,
Nucl.\ Phys.\ B301 (1988) 554}]. By fitting fig.\ 8 of this reference
we obtain:
$$
n_{\rm ch}(W) = 1.67 + 0.211 \,\, {\rm exp} \left[ 3.06
\, \ln^{1/2} \left( { W \over {\rm GeV} } \right) \right] .
\eqn \avi
$$
We assume now that all charged particles in the first generation of the shower
are above the \v{C}erenkov
threshold. This is a good
approximation at the high energies that will turn out to be relevant
for hadronic blasts. With this assumption, $X_d(W)$ is the product of $
n_{\rm ch}(W) $ and the mean path length of a charged particle. A lower
bound to the latter is one nuclear interaction length, $X_{\rm
nucl} = 84.9 $ g/cm${}^2$ in water. This leads to a lower bound for
$X_d(W)$:
$$
X_d(W) > X_{\rm min}(W) \simeq n_{\rm ch}(W) X_{\rm nucl} .
\eqn \avii
$$
An upper bound is obtained by multiplying $X_{\rm
min}(W)$ by the typical length of a hadronic shower in units of
interaction lengths [\Ref\lambdaI{Ref.~[\revpartprop] [section
III.23]}]:
$$
X_d(W) < X_{\rm max}(W) \simeq n_{\rm ch}(W) X_{\rm nucl}
\left[ 5.45 + 0.89 \ln \left( { W \over {\rm GeV}} \right) \right] .
\eqn \aviii
$$
A lower and an upper bound to the visible energy $E_{\rm vis}(W)$ can then
be obtained from eq.~\av.

This range then has to be compared with the IMB energy threshold for
$\pi^\pm$ detection, $E_{\pi} = 1.52 \, m_{\pi} + 30 \, {\rm MeV} + 140
\, {\rm MeV} = 382 \, {\rm MeV}$, and with the highest energy analyzed,
$E_{\rm vis} = 2500 \,{\rm MeV}$ [\IMBc]. The result of such a
comparison is that for $W \simle 10^2 \, {\rm TeV}$ there is probably
not enough \v{C}erenkov light for the `blast' to be seen, while for $W
\simge 10^6 \, {\rm TeV}$ the detector is probably overloaded (more
than 900 PMTs fired [\IMBc]) or otherwise not sufficiently efficient to
detect the blast. However for $10^2 \, {\rm TeV} \simle W \simle 10^6
\, {\rm TeV}$, such blasts, if they do occur, should already be present
in the sample of ref.~[\IMBc], probably as multiple-ring events.

However, the neutrino energy $E_\nu$ required to have $W \simge 10^2 \,
{\rm TeV}$ is $E_\nu \simge 10^7 \,\TeV$. This comes  from the
kinematic relation $ W^2 \simeq 2 m_p E_\nu ( 1 - x ) y$, with $m_p$
the nucleon mass and $x$ and $y$ the usual deep inelastic scattering
variables, together with the consideration that at high energies the
$\W$ or $\Z$ propagator restricts the important values of $x$ and $y$ to
$x \simeq 0$ and $y \simeq 1$. So `hadronic blasts' in \v{C}erenkov
detectors  turn out to be important only for very high energy
neutrinos, where (at least for the purposes of this paper) there
already are much better bounds on cosmic neutrino fluxes from EAS arrays.

For the sake of completeness, we show in fig.~\ffour a (curve 1 at high
energies) the contribution of hadronic blasts in IMB, obtained by numerical
integration of:
$$
P_{\rm hadr}^\cont(E_\nu) = { N_{\rm nucl} \over M} \int_0^{2m_pE_\nu}
{\rm d}\,Q^2 \int_{W^2_{\rm min}}^{W^2_{\rm max}} {\rm d}\,W^2 {1 \over
(2m_pE_\nu)^2 y} \left[ {{\rm d}\, \sigma^{CC} \over {\rm d}\,x \, {\rm
d}\,y } + {{\rm d}\, \sigma^{NC} \over {\rm d}\,x \, {\rm d}\,y }
\right].
\eqn \aix
$$
Here $Q^2\simeq 2m_pE_\nu xy$, $W_{\rm min} = 10^2 \TeV$,
$W_{\rm max} = 10^6 \TeV$ and the contributions from charged and neutral
currents have been summed.  The EHLQ structure functions have been used
together with a McKay-Ralston asymptotic form at low $x$ (as in
ref.~[\RenoQuigg]).

\vfill \eject
\refout

\vfill \eject
\figout

\bye